\documentclass[epj]{webofc}
\usepackage[varg]{txfonts}   
\woctitle{MENU 2013}
\begin{document}
\title{Beyond Rainbow-Ladder in a covariant three-body Bethe-Salpeter approach: Baryons}
%
%

\author{H\`elios Sanchis-Alepuz\inst{1}\fnsep\thanks{\email{Helios.Sanchis-Alepuz@theo.physik.uni-giessen.de}}\footnote{speaker}
\and
Stanislav D. Kubrak\inst{1}\fnsep\thanks{\email{Stanislav.D.Kubrak@physik.uni-giessen.de}}
\and
Christian S. Fischer\inst{1}\fnsep\thanks{\email{Christian.Fischer@theo.physik.uni-giessen.de}}
}

\institute{Institut f\"ur Theoretische Physik, Justus-Liebig Universit\"at Giessen, Giessen, Germany}

\abstract{%
We report on recent results of a calculation of the nucleon and delta masses in a covariant 
bound-state approach, where to the simple rainbow-ladder gluon-exchange interaction kernel 
we add a pion-exchange contribution to account for pion cloud effects. We observe good
agreement with lattice data at large pion masses. At the physical point our masses are 
too large by about five percent, signaling the need for more structure in the gluon part of 
the interaction.
}
\maketitle
\section{Introduction}
\label{intro}

Continuum functional methods, when applied to hadron phenomenology, have as their ultimate goal, the calculation of hadronic properties using QCD's elementary degrees of freedom. Mesons and baryons are thus considered as bound states of quarks and, hence, described by two- and three-body Bethe-Salpeter equations (BSEs). The study of these equations relies upon the knowledge of several QCD's Green's functions which can, in turn, by obtained as solutions of Dyson-Schwinger equations (DSEs). The approach is Poincar\'e covariant and is applicable at any momentum range.

However, in practice, truncations of both the DSEs and the bound state equations are necessary. The simplest one consistent with Poincar\'e covariance and with chiral symmetry is the Rainbow-Ladder truncation (RL). This truncation has been extensively and successfully used in ground-state meson and baryon calculations (see e.g. \cite{Eichmann:2013afa,Bashir:2012fs} for overviews).

On the other hand, much work has been 
invested in the past years on the extension of RL towards more 
advanced approximations of the quark-gluon interaction by improving our knowledge of the
quark-gluon vertex 
\cite{Fischer:2005en,Chang:2009zb,Chang:2010hb,heupel_new}.
Another promising, and more accessible, approach consists on parametrizing
important unquenching effects in the quark-gluon interaction by the inclusion of hadronic degrees 
of freedom \cite{Fischer:2007ze,Fischer:2008sp,Fischer:2008wy}. 
This is possible, since the vertex DSE can be decomposed 
diagrammatically into contributions stemming from Yang-Mills theory
and those involving quark-loops. To leading order in the bound state
mass, the latter ones can be expressed in terms of (off-shell) 
pion exchange between quarks.

In these proceedings we report on our recent results for the spectrum of light baryons when pion-cloud effects are included. A more detailed description of the calculation as well as a thourough analysis of the results herein can be found in \cite{Sanchis-Alepuz:2014wea}. 

\section{Covariant Faddeev equation}
\label{sec-1}

\begin{figure}
\centering
\includegraphics[width=0.9\textwidth,clip]{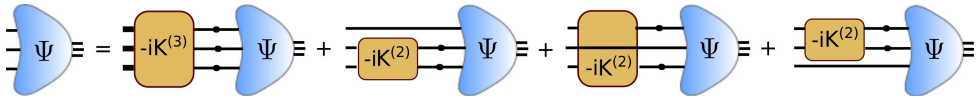}
\caption{Schematic representation of the three-body Bethe-Salpeter equation.}
\label{fig-1}       
\end{figure}

The mass and internal structure of baryons are given, in a covariant
Faddeev approach, by the solutions of the three-body equation
\begin{equation}\label{eq:3bBSEcompact}
\Psi = -i\widetilde{K}^{(3)}~G_0^{(3)}~\Psi + \sum_{a=1}^3 -i\widetilde{K}_{(a)}^{(2)}~G_0^{(3)}~\Psi\,,
\end{equation}
where $\widetilde{K}^{(3)}$ and $\widetilde{K}^{(2)}$ are the three- and two-body interaction kernels, respectively, and $G_0$ represents the product of three fully-dressed quark propagators $S$. This equation is depicted in Fig.\ref{fig-1}.
The amplitude $\Psi$ contains all the information about the state. In particular, its spin-momentum part can be decomposed into partial wave sectors and Poincar\'e covariance enforces that it contains all spin and orbital angular momentum contributions allowed by angular momentum addition rules.

The quark propagators are obtained from their respective DSE
\begin{equation}\label{eq:quarkDSE}
 S^{-1}(p)=S^{-1}_0(p)+Z_{1f}\int_q \Gamma^\nu_{gqq,0}
D_{\mu\nu}(p-q)\Gamma^\nu_{gqq}(p,q)S(q)\,\,,
\end{equation}
where $S_0$ is the (renormalized) 
bare propagator, $\Gamma^\nu_{gqq}$ is the full quark-gluon vertex with 
its bare counterpart $\Gamma^\nu_{gqq,0}$, $D^{\mu\nu}$ is the full
gluon propagator and $Z_{1f}$ and $Z_2$ are renormalization constants.

\begin{figure}
\centering
\includegraphics[width=0.5\textwidth,clip]{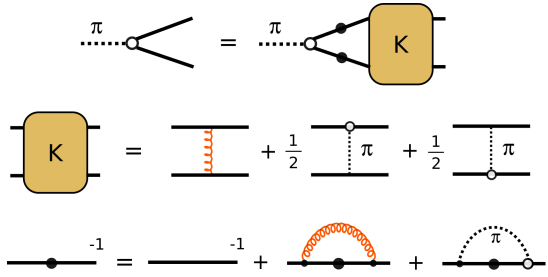}
\caption{Schematic representation of the pion BSE, two-body kernel and quark DSE when pion-exchange contributions are included. Big and small blobs represent full and bare vertices, respectively.}
\label{fig-2}       
\end{figure}

At the level of the BSE, the RL truncation  consists of reducing the 2-body kernel to a single dressed-gluon exchange, which couples to the quarks via a dressed vector-vector interaction, and neglecting the 3-body kernel altogether. For the DSE, this truncation corresponds to taking only the vector part of the quark-gluon vertex, multiplied by the same dressing function used in the BSE kernel. In this work, we extend this truncation the exchange of a pion which couples to quarks through the leading component of its BSE amplitudes, that is, via a dressed $\gamma_5$ vertex; see Fig.\ref{fig-2}

\section{Results}

The dressing function multiplying the quark-gluon vertex in the RL truncation must be modelled. In this work we use the model introduced in \cite{Maris:1997tm,Maris:1999nt}. Its UV part simply reproduces the one-loop perturbation theory behaviour of the strong coupling, whereas the IR part is designed to enable dynamical chiral-symmetry breaking and depends on two parameters $\eta$ and $\Lambda$. These, together with the current-quark mass, are fit to reproduce the physical pion mass and decay constant. These are largely insensitive to the value of $\eta$ between $\eta=1.6$ and $\eta=2.0$. The dependence of our results on the $\eta$-value is taken here as an indication of the model dependence of the calculation. The value of $\Lambda$ depends on whether only the RL kernel is used or also the pion exchange is included. Its values are $\Lambda=0.74$~GeV in the first case (which we call RL1) $\Lambda=0.84$~GeV in the second case (RL2). In both cases, the current-quark mass is $m_{u/d}(19GeV)=3.7$~MeV.

\begin{figure}[b]
\centering
\includegraphics[width=0.8\textwidth,clip]{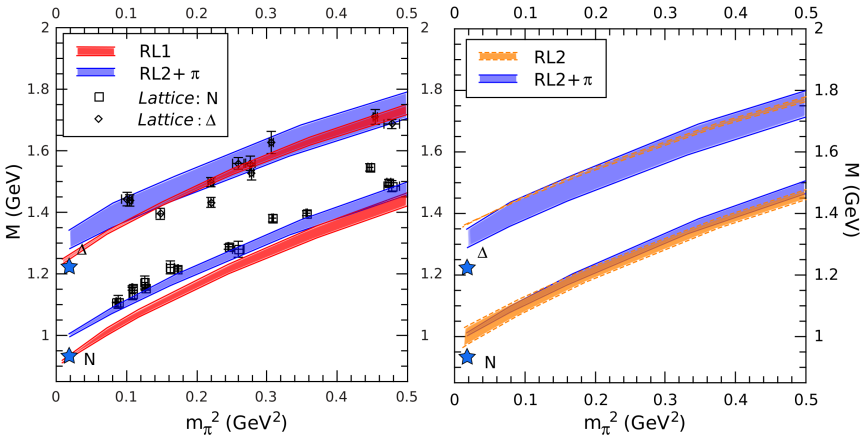}
\caption{Evolution of the nucleon and delta mass with respect to the pion mass squared. \textit{Left panel}: We plot the results for pure RL1 and for RL2 with pion exchange. We also compare with a selection of lattice data \cite{Ali Khan:2003cu}-\cite{Gattringer:2008vj}. \textit{Right panel}: We compare the results for RL2 only and RL2 with pion exchange. Stars denote the physical nucleon and delta mass.
The shaded bands correspond to a variation of the interaction parameter $\eta$ between
$1.6 \le \eta \le 2.0$, with $\eta=1.6$ corresponding to the upper limit of the bands.}
\label{fig-3}       
\end{figure}

In Fig.\ref{fig-3} we display
the baryon-mass evolution with the squared pion mass (or, equivalently, with 
respect to the current-quark mass). 
The comparison with lattice data suggests that the binding provided by the pure
rainbow-ladder scheme RL1 is somewhat too strong at larger pion masses, although the agreement with the experimental nucleon and delta masses is excellent. Taking the pion back-reaction into account we find now good
agreement with the lattice results even down to the lattice points with lowest
pion mass. However, it is also clear that the curvature of $M(m_\pi^2)$ below
this point is slightly too small. As a result, our value of the nucleon and delta
masses at the physical point are slightly too large. It will be interesting to see,
whether this can be remedied by the inclusion of further structure in the  
quark-gluon interaction beyond the present rainbow-ladder scheme. 


\section*{Acknowledgments}
This work was supported by the Helmholtz 
International Center for FAIR within the LOEWE program of the State of 
Hesse, by the Helmholtz Center GSI, by the Erwin Schr\"odinger 
fellowship J3392-N20 of the FWF and by the DFG transregio TR 16.

%
%

\end{document}